\documentclass[aps,prb,twocolumn,showpacs, superscriptaddress]{revtex4}

\usepackage{amsmath,amssymb}
\usepackage{graphicx}
\usepackage{bm}
\usepackage{epstopdf}
\usepackage{color}

\newcommand{\laf}{LaFeAsO$_{1-x}$F$_{x}$}

\newcommand{\lafour}{LaFeAsO$_{0.96}$F$_{0.04}$}
\newcommand{\LaOFxFeAs}{LaFeAsO$\mathrm{_{1-x}}$F$\mathrm{_{x}}$}
\newcommand{\bfca}{Ba(Fe$_{1-x}$Co$_x$)$_2$As$_2$}
\newcommand{\smaf}{SmFeAsO$\mathrm{_{1-x}}$F$\mathrm{_{x}}$}
\newcommand{\Ceaf}{CeFeAsO$\mathrm{_{1-x}}$F$\mathrm{_{x}}$}
\newcommand{\la}        {$^{139}$La}

\newcommand{\slrrt}     {$(T_1T)^{-1}$}
\newcommand{\slrr}      {$T_1^{-1}$}

\begin{document}

\title{Progressive slowing down of spin fluctuations in underdoped \LaOFxFeAs}

\author{F. Hammerath}
\email[]{franziska.hammerath@unipv.it}
\affiliation{Dipartimento di Fisica and Unit\`a CNISM di Pavia, I-27100 Pavia, Italy}
\affiliation{IFW Dresden, Institute for Solid State Research, PF
270116, 01171 Dresden, Germany}
\author{U. Gr\"afe}
\affiliation{IFW Dresden, Institute for Solid State Research, PF
270116, 01171 Dresden, Germany}
\author{H. K\"uhne}
\affiliation{National High Magnetic Field Laboratory, Florida State University, 
Tallahassee, Florida 32310, USA}
\author{P.~L. Kuhns}
\affiliation{National High Magnetic Field Laboratory, Florida State University, 
Tallahassee, Florida 32310, USA}
\author{A.~P. Reyes}
\affiliation{National High Magnetic Field Laboratory, Florida State University, 
Tallahassee, Florida 32310, USA}
\author{G. Lang}
\affiliation{IFW Dresden, Institute for Solid State Research, PF
270116, 01171 Dresden, Germany}
\affiliation{LPEM UMR8213, CNRS - ESPCI ParisTech - UPMC, 75005 Paris, France}
\author{S. Wurmehl}
\affiliation{IFW Dresden, Institute for Solid State Research,
PF 270116, 01171 Dresden, Germany}
\affiliation{Institute for Solid State Physics, Dresden Technical University, TU-Dresden, 01062 Dresden, Germany}
\author{B. B\"{u}chner}
\affiliation{IFW Dresden, Institute for Solid State Research,
PF 270116, 01171 Dresden, Germany}
\affiliation{Institute for Solid State Physics, Dresden Technical University, TU-Dresden, 01062 Dresden, Germany}
\author{P. Carretta}
\affiliation{Dipartimento di Fisica and Unit\`a CNISM di Pavia, I-27100 Pavia, Italy}
\author{H.-J. Grafe}
\affiliation{IFW Dresden, Institute for Solid State Research, PF
270116, 01171 Dresden, Germany}
\date{\today}

\begin{abstract}

The evolution of low-energy spin dynamics in the iron-based
superconductor \laf\ was studied over a broad doping, temperature, and magnetic field range ($x$ = 0 -- 0.15, $T \leq$ 480\,K, $\mu_0H \leq$ 30\,T) by means of
$^{75}$As nuclear magnetic resonance (NMR). An enhanced
spin-lattice relaxation rate divided by temperature,
$(T_1T)^{-1}$, in underdoped superconducting samples ($x$ = 0.045,
0.05 and 0.075) suggests the presence of antiferromagnetic spin
fluctuations, which are strongly reduced in optimally-doped
($x=0.10$) and completely absent in overdoped ($x=0.15$) samples.
In contrast to previous analysis, Curie-Weiss fits are shown to be
insufficient to describe the data over the whole temperature
range. Instead, a BPP-type model is used to describe the
occurrence of a peak in $(T_1T)^{-1}$ clearly above the
superconducting transition, reflecting a progressive slowing down of the spin fluctuations down to the superconducting phase transition.

\end{abstract}


\maketitle

\section{Introduction}

Investigations of the role of spin fluctuations in iron-based
superconductors are crucial for the understanding of the mechanism
of superconductivity in these compounds. Standard electron-phonon
modes have been found to be too weak to mediate superconductivity
with the reported transition temperatures.\cite{Boeri2008,
Mazin2008} Instead, the vicinity to a magnetically-ordered ground
state and the topology of the multiband Fermi surface with
quasi-nested electron and hole pockets triggered 
already at a very early stage of the pnictide research era theoretical considerations that spin
fluctuations might be the pairing glue for Cooper
pairs, \cite{Mazin2008, ChubukovPRB2008, KurokiPRL2008,Mazin2009} similar to the previous case of cuprate superconductors.

Nuclear Magnetic Resonance (NMR) is a versatile local probe
technique to study the superconductivity as well as the static and
dynamic magnetic properties of a material. The NMR spin-lattice
relaxation rate, $T_1^{-1}$, is a very useful probe of the magnetic
fluctuations, since it is directly proportional to the wave-vector
$\vec{q}$ dependent dynamic spin susceptibility
$\chi''(\vec{q},\omega)$. It is thus a key tool to
investigate spin fluctuations on the border of magnetism and
superconductivity in the iron-based superconductors. Indeed, an
enhanced nuclear spin-lattice relaxation rate divided by
temperature, \slrrt , has been observed in a number of non-magnetic,
superconducting iron pnictides, indicating the existence of strong
antiferromagnetic spin
fluctuations.\cite{Ning2009,Imai2009,NakaiPRL2010,Ning2010,Kawasaki2010,Kinouchi2011,Oka2012,NakaiJPSJ2008,NakaiNJP2009,Nakano2010}
However, the role of these fluctuations for the occurrence of
superconductivity is heavily debated. Some references still find
pronounced spin fluctuations in optimally-doped samples with the
highest
$T_c$,\cite{Ning2009,Imai2009,Ning2010,NakaiPRL2010,Kawasaki2010,Kinouchi2011,Oka2012}
concluding that these fluctuations promote superconductivity.
Other references report that the highest $T_c$ correlates with the
complete suppression of previously existing spin fluctuations,
which rather points towards a competition of superconductivity and
magnetism.\cite{GrafePRL2008,NakaiJPSJ2008,MukudaJPSJ2009,NakaiNJP2009,Nakano2010}

Here, we study the evolution of spin fluctuations as a function of doping, temperature, and magnetic field in \laf, where superconductivity in the
FeAs-layers arises upon substituting oxygen by fluorine in the
LaO-layer.\cite{Kamihara2008} This out-of-plane electron-doping is
expected to minimize the influence of the dopants on the FeAs
planes, in contrast to in-plane Cobalt dopants for example, which 
additionally act as impurity
scatterers.\cite{Wadati2010} Furthermore, the transition from the
magnetically-ordered state to the superconducting one is abrupt in
\laf.\cite{LuetkensNatMat2009, Oka2012} No sign of coexistence of
superconductivity and static magnetism has been observed in this
compound, which is in contrast to other pnictide families such as \Ceaf, \smaf\ or Co-doped \bfca.
\cite{Drew2009, Sanna2010, Laplace2009} This exclusive superconducting or magnetic ordering enables to study the role of fluctuations without the additional complication of contributions from reminiscent magnetic order. Our measurements are largely consistent with previous
NMR investigations of \laf, which reported the absence of
antiferromagnetic spin fluctuations for optimally-doped ($x=0.1$)
and overdoped samples,\cite{GrafePRL2008,NakaiJPSJ2008} and the
presence of such fluctuations in underdoped
samples,\cite{NakaiJPSJ2008, NakaiNJP2009,Nakano2010} pointing
towards a competition of magnetism and superconductivity. We
extend these previous investigations to higher temperatures,
higher fields and a broader range of doping levels.  
Note that a recent nuclear quadrupole resonance (NQR) study
suggested a slightly different phase diagram, where significant spin fluctuations still
appear in optimally-doped (in this case $x=0.06$)
samples.\cite{Oka2012}

This report focuses on the presentation of a new approach to quantitatively describe the relaxation data of underdoped superconducting, non-magnetic samples. To date, all published analyses of the enhancement of $(T_1T)^{-1}$ with
decreasing temperature in such samples were based on a Curie-Weiss picture
\cite{NakaiJPSJ2008,NakaiPRL2010,Kawasaki2010,Nakano2010,Kinouchi2011,Oka2012}
or a combination of a Curie-Weiss term and an activated temperature dependence, to
account for the decrease of $(T_1T)^{-1}$ with decreasing
temperature in the high-temperature regime.\cite{Ning2010} 
The common use of the Curie-Weiss model for the increase of \slrrt\ follows from Moriya's self-consistent renormalization (SCR) theory for weakly itinerant two-dimensional (2D) antiferromagnets, which showed that the staggered susceptibility above $T_N$ can be approximately described by a Curie-Weiss law.\cite{Hasegawa1974, Ueda1975}
However, in some of these studies there is a noticeable deviation
from the Curie-Weiss law at low temperatures: $(T_1T)^{-1}$
decreases visibly already above the superconducting transition and
forms a well-defined peak above
$T_c$,\cite{NakaiJPSJ2008,Imai2009} which can neither be expressed
within the Curie-Weiss model nor within the more correct SCR framework. We focus on the appearance of this peak, which 
is present for all our underdoped samples and
propose an analysis based on the
Bloembergen-Purcell-Pound (BPP) model,\cite{BloembergenNature1947,
Bloembergen1948} which can be used to describe the progressive slowing
down of spin fluctuations. Our studies of the doping and field
dependence of the peak in $(T_1T)^{-1}$ as well as selected
measurements of the spin-spin relaxation rate, $T_2^{-1}$,
corroborate the choice of the BPP-model and help to rule out other possible origins of such a peak in \slrrt, such 
as spin diffusion effects,\cite{Kambe1994, Julien2008} or a field-induced anisotropy in the spin fluctuations.\cite{Suh1995}

\section{Sample preparation and Experimental details}
\label{sampleprep}

Polycrystalline samples of \laf\ with nominal doping levels $x$ = 0, 0.035, 0.045, 0.05, 0.075, 0.1 and 0.15 have been prepared by following and improving the two-step solid state reaction approach of Zhu {\it et al.} \cite{Zhu2008, Kondrat2009} Detailed structural, thermodynamic and transport characterization studies on the samples with $x$ = 0, 0.05, 0.075, 0.1 and 0.15 can be found in previous publications.\cite{Klauss2008a,Luetkens2008, LuetkensNatMat2009, Hess2009, Kondrat2009, Klingeler2010}
The undoped sample ($x = 0$) shows a structural transition at $T_s$ = 156\,K, followed by a magnetic ordering at $T_N$ = 138\,K.
Samples with $x > 0.04$ are superconducting with $T_c$ = 20/22/26.8/10 K for $x$ = 0.05/0.075/0.1/0.15. The presence of static magnetic order in these superconducting samples has been ruled out experimentally.\cite{Luetkens2008, LuetkensNatMat2009}
Two new samples with $x$ = 0.035 and $x$ = 0.045, residing directly at the boundary between the magnetically-ordered and the superconducting ground state have been prepared recently. The sample with $x$ = 0.035 is not superconducting. The temperature dependence of its susceptibility resembles that of other magnetically-ordered samples with $x < 0.04$.\cite{Klingeler2010} Very slight changes of slope in the region between 80 and 140\,K indicate possible structural and magnetic transitions around $T_s \approx 120$\,K and $T_N \approx 100$\,K. 
However, these anomalies in the susceptibility are too weak to determine the exact values of $T_s$ and $T_N$. We will show that NMR $T_1$ measurements on this sample are able to determine the magnetic transition very precisely. The sample with $x$ = 0.045 on the other hand does not exhibit any magnetic or structural ordering, but is superconducting with $T_c \approx$ 19\,K, as probed by magnetization measurements in a magnetic field of 20\,Oe. 

For NMR measurements, the pellets were ground to a powder of 1--100 $\mu$m grain size. To protect the samples from moisture, they were put into quartz glass tubes which were sealed with Teflon thread tape.

$^{75}$As (nuclear spin $I=3/2$) NMR measurements were carried
out in a magnetic field of 7\,T for all samples and additionally in 16\,T for the sample with $x$ = 0.035 and in 3\,T, 16\,T, 23\,T, and 30\,T for the sample with $x$ = 0.045. Since the superconducting transition temperature decreases in an applied magnetic field and the knowledge of $T_c(H)$ was crucial for our analysis, we determined $T_c(H)$ by in situ ac susceptibility measurements by tracking the detuning of the NMR resonance circuit. The corresponding values for $T_c$(7\,T) for all doping levels and $T_c(H)$ for the sample with $x=0.045$ can be found in the first row of Tables \ref{tab:Tc} and \ref{tab:Tc4.5}, respectively. Due to the way $T_c$ was determined, it should be noted that for $\mu_0H>0$ some corrections to $T_c(H)$ can be present, due to the onset of vortex motions, which will however not impair our analysis.
 The nuclear spin-lattice relaxation rate, $T_1^{-1}$, was
measured on the high-frequency peak of the quadrupolar-broadened NMR powder pattern of the central transition ($I_z = 1/2 \to I_z = -1/2$).\cite{GrafePRL2008} This peak corresponds to the field orientation $H || ab$, i.e., to the direction parallel to the iron planes.\cite{GrafePRL2008} The exact position of the peak was determined via frequency scans for each temperature, to exclude possible frequency-dependent effects on \slrr. The inversion recovery method was used to determine $T_1$ and the recovery of the nuclear magnetization was fitted to the relaxation formula for the central transition of a nuclear spin $I = 3/2$:
\begin{equation}
M_z(t)=M_0\left[1-f\left(0.9\rm{e}^{(-(6t/T_1)^{\lambda})}+0.1\rm{e}^{(-(t/T_1)^{\lambda})}\right)\right] \, 
\label{eq:T1}
\end{equation}
where $M_z(t)$ is the nuclear magnetization recovered after a certain time $t$, $M_0$ is the saturation magnetization at thermal equilibrium, $f$ is the inversion factor, which for a complete inversion equals 2, and $T_1$ is the nuclear spin-lattice relaxation time.
A stretching exponent $\lambda$ with $1>\lambda \geq 0.45$ had to be used below 100\,K to fit the recovery curves of all samples except for $x=0$, $x=0.1$, and $x=0.15$. In these three cases, $\lambda$ could be kept to 1 for all temperatures.

\section{Experimental Results and Analysis}

\begin{figure}[]
\begin{center}
\includegraphics[width=\columnwidth,clip]{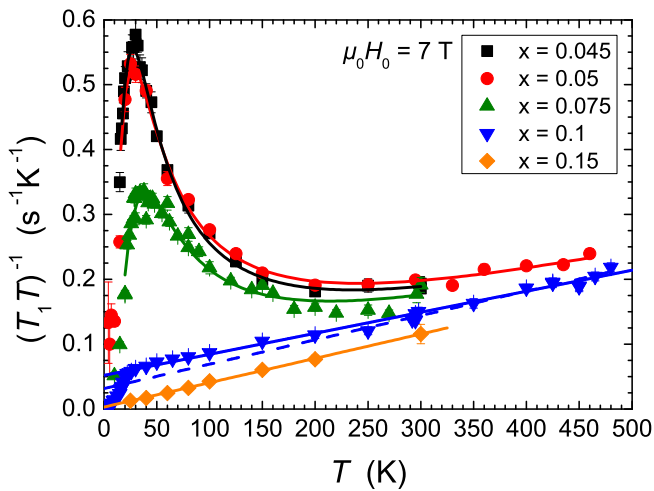}
\caption{\label{fig:dop} (color online) The spin-lattice relaxation rate
divided by temperature, \slrrt , versus temperature, measured in a magnetic field of 7\,T, for different
doping levels of superconducting samples. The lines are fits to our proposed BPP-like model Eq.~\eqref{eq:BPP} for $0.045 \leq x \leq 0.075$ and linear fits for $x= 0.1$ and $x=0.15$ (solid and dashed lines, see text for details).}
\end{center}
\end{figure}

Fig.~\ref{fig:dop} shows the temperature evolution of the
$^{75}$As NMR spin-lattice relaxation rate divided by temperature,
$(T_1T)^{-1}$, measured in a magnetic field of 7\,T, for all
superconducting samples ($x\geq 0.045$). For selected samples, the
measurements have been extended up to 480\,K ($x = 0.05$ and
$x=0.1$).

\begin{table}[b]
    \center
    \caption{$T_c$ from in situ ac susceptibility measurements in 7\,T and $T_{max}$ of \slrrt\  for all investigated doping levels.}
    \begin{tabular}{|c|c|c|c|c|c|}
     \hline
     $x$ & 4.5 \% & 5\% & 7.5\% & 10\% & 15\%\\
     \hline
     \hline
     $T_c$ & 16 K & 16 K & 18 K & 22 K & 9 K\\
     \hline
     $T_{max}$ & 27 K & 28 K & 40 K & - & -\\
     \hline
   \end{tabular}
   \label{tab:Tc}
\end{table}

\begin{figure}[t]
\centering
\includegraphics[width=\linewidth]{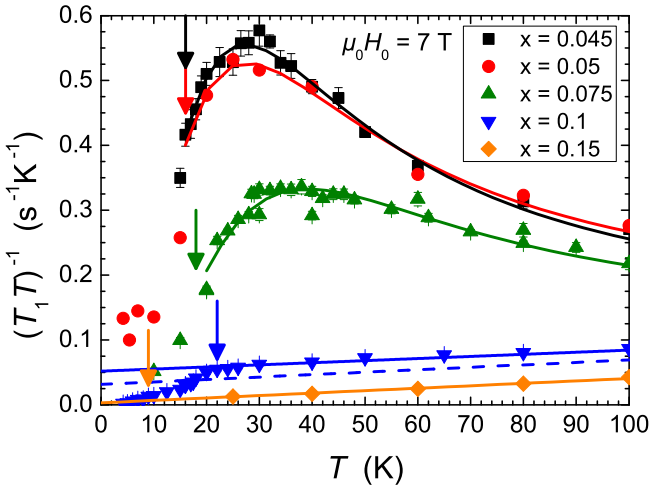}
\caption{\label{fig:zoom} (color online) Enlargement of the low
temperature region of Fig.~\ref{fig:dop}. Lines are fits to the data (symbols) with Eq.~\eqref{eq:BPP} for $0.045 \leq x \leq 0.075$ and linear fits for $x= 0.1$ and $x=0.15$ (solid and dashed lines, see text for details). The superconducting transition
temperatures, $T_c$(7\,T), are marked by arrows.}
\end{figure}

For the optimally-doped ($x=0.1$) and the overdoped ($x=0.15$)
samples, $(T_1T)^{-1}$ decreases monotonically with decreasing
temperature. This behavior is known since the beginning of the
research on iron-based superconductors. At this early stage, it had
been compared to the pseudogap behavior in the
cuprates.\cite{GrafePRL2008,NakaiJPSJ2008} The NMR Knight shift,
which is a direct measure of the intrinsic static spin
susceptibility, $\chi(\vec{q}=0, \omega=0)$, shows a similar temperature
dependence.\cite{GrafePRL2008, Imai2008} However, no pseudogap peak could be observed up to
480\,K and activated fits, as usually used to describe the opening of a pseudogap, fail to describe the data consistently
over the whole temperature range, which renders the pseudogap
scenario rather unlikely.\cite{GrafeNJP2009,PaarPhysicaC2010} In
contrast to the previously intended pseudogap fits, \slrrt\ of $x=
0.1$ and $x= 0.15$ can be well fit with a simple linear
temperature dependence over the whole temperature range down to
$T_c$ (see figures~\ref{fig:dop} and~\ref{fig:zoom}). This agrees well with the static
susceptibility measured by SQUID magnetization measurements, which
also shows a linear temperature dependence in the high-temperature regime.\cite{Klingeler2010} Several theoretical
approaches have been made to discuss the linear decrease of the
static uniform susceptibility with decreasing temperature,
including the consideration of antiferromagnetic
fluctuations\cite{KorshunovEPL2008,KorshunovPRL2009,ZhangEPL2009}
and of a large polarizability of the anions leading to attractive
excitonic interactions and possibly to the preformation of Cooper
pairs well above the superconducting
transition.\cite{Berciu2009,Sawatzky2009} A recent theoretical
paper suggests peculiarities in the orbitally-resolved density of
states to be the reason for the decreasing
susceptibility.\cite{Skornyakov2011} Another recent investigation suggests that
average effective local iron spins, $S_{eff}$, result from a dynamical mixing of different iron spin states and that singlet correlations among these $S_{eff}$ are causing the peculiar temperature dependence of the susceptibility.\cite{Chaloupka2013}

In the underdoped samples ($x$ = 0.045, 0.05, and 0.075), at high temperatures, $(T_1T)^{-1}$
shows a similar decrease with decreasing temperature as the
optimally- and over-doped samples. But below 200\,K $(T_1T)^{-1}$
increases, indicating the presence of pronounced antiferromagnetic
spin fluctuations. These fluctuations are only observable in
$(T_1T)^{-1}$ and not in the macroscopic spin susceptibility or
the NMR Knight shift.\cite{GrafeNJP2009,Klingeler2010} This is
because $(T_1T)^{-1}$ probes the $\vec{q}$-integrated imaginary part
of the dynamical spin susceptibility, $\chi''(q,\omega)$, whereas the macroscopic spin susceptibility or the
Knight shift probe only at $\vec{q}=0$.
Upon increasing the doping-level, the increase of \slrrt\ is reduced. This behavior is quite common for underdoped 
samples of \LaOFxFeAs\ and other pnictide families. \cite{Ning2009,Imai2009,NakaiPRL2010,Ning2010,Kawasaki2010,Kinouchi2011,Oka2012,NakaiJPSJ2008,NakaiNJP2009,Nakano2010}

\begin{figure}
\centering
\includegraphics[width=\linewidth]{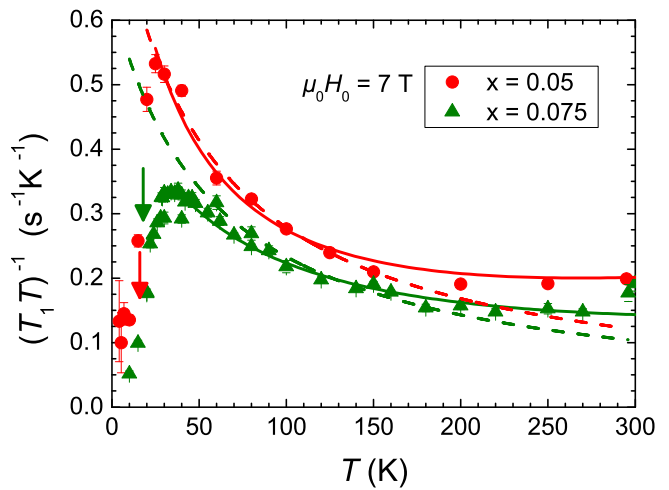}
\caption{\label{fig:CW} (color online) Examples of Curie-Weiss fits (dashed lines) and fits including a Curie-Weiss contribution and a linear temperature dependence (solid lines) to the $(T_1T)^{-1}$ data of the samples with $x=0.05$ (red dots) and $x=0.075$ (green triangles).
The superconducting transition temperatures, $T_c$(7\,T), are marked by arrows.}
\end{figure}

Fig.~\ref{fig:CW} shows Curie-Weiss fits of the form $(T_1T)^{-1}
= C/(T+\theta)$ (dashed lines) for $x=0.05$ and
$x=0.075$. Such fits have been widely used to describe the
increase of spin fluctuations towards
$T_c$.\cite{NakaiJPSJ2008,NakaiPRL2010,Kawasaki2010,Nakano2010,Kinouchi2011,Oka2012}
However, as can be seen in Fig.~\ref{fig:CW}, the fits are unable to describe the data for $T>200$\,K and at low
temperatures. The deviations at high temperatures can be addressed by adding a linear temperature-dependence of \slrrt, as
used for $x=0.1$ and $x=0.15$ (see solid lines in Fig.~\ref{fig:CW}). However, much more important is the
fact that $(T_1T)^{-1}$ decreases already above $T_c$ and
apparently forms a well-defined peak. This peak is visible in the
$(T_1T)^{-1}$ data of all underdoped samples (see also
Fig.~\ref{fig:zoom}) and it cannot be described with the depicted Curie-Weiss fit combinations. Table \ref{tab:Tc} compares the temperature
of the maximum of $(T_1T)^{-1}$, $T_{max}$, to the corresponding
$T_c$ in the same field, measured by in situ ac susceptibility
measurements. The maximum of $(T_1T)^{-1}$ occurs well above
$T_c$. Such a peak has already been observed in other iron-based
(non-magnetic) superconductors, such as underdoped \lafour\ and
FeSe under pressure.\cite{NakaiNJP2009, Imai2009} It has been
interpreted as a weak magnetic ordering,\cite{NakaiNJP2009} which does not appear to be compatible with
our data in light of previous experimental evidence,\cite{Oka2012, LuetkensNatMat2009} or as a
glassy spin freezing,\cite{Imai2009} whose detailed analysis remains to be done.

\begin{table}[b]
    \center
    \caption{$T_c(H)$ of the sample with $x=0.045$ from in situ ac susceptibility measurements in the different applied magnetic fields and $T_{max}(H)$ of \slrrt\ for the  same sample.}
    \begin{tabular}{|c|c|c|c|c|c|c|}
     \hline
      $\mu_0H$& 0 T & 3 T & 7 T & 16 T & 23 T & 30 T\\
     \hline
     \hline
     $T_c$ & 20 K & 18 K & 16 K & 15 K & 12 K & 10 K \\
     \hline
     $T_{max}$ & - & 25 K & 27 K & 36 K & 38 K &  42 K \\
     \hline
   \end{tabular}
   \label{tab:Tc4.5}
\end{table}

\begin{figure}
\centering
\includegraphics[width=\linewidth]{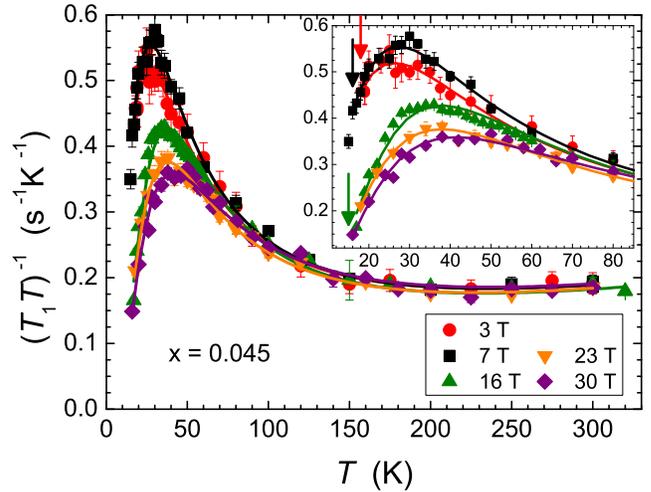}
\caption{\label{fig:field} (color online) Temperature-dependent \slrrt\ of
LaFeAsO$\mathrm{_{0.955}}$F$\mathrm{_{0.045}}$, measured in magnetic fields of 3\,T (red dots), 7\,T (black squares), 16\,T (green triangles), 23\,T, (orange triangles) and 30\,T (purple diamonds). Solid lines are fits
to our proposed BPP-like model, see Eq.~\eqref{eq:BPP}. The inset depicts the relevant low-temperature region. Arrows mark $T_c(H)$.}
\end{figure}

Insight on the proper description of the relaxation data can be gained by measuring the evolution of this peak upon changing the external magnetic field. Measurements of \slrrt\ on an underdoped sample with $x$ = 0.045 in magnetic fields of 3, 7, 16, 23, and 30\,T are shown in Fig.~\ref{fig:field}. While the high-temperature behavior of the spin-lattice relaxation rate remains unaffected by the application of higher fields, the position and the height of the peak in \slrrt\ clearly change with field. Mainly, the temperature $T_{max}$ where \slrrt\ is maximal shifts to higher temperatures (from 25\,K in 3\,T to 42\,K in 30\,T) and thus is even more clearly above the superconducting transition temperature $T_c$, which is itself reduced by the application of the external magnetic field, as expected (see also Table~\ref{tab:Tc4.5}).

The field dependence of this peak is reminiscent of the characteristic field dependence of the BPP (Bloembergen-Purcell-Pound) model,\cite{BloembergenNature1947,Bloembergen1948}
which describes the behavior of the spin-lattice relaxation rate, $T_1^{-1}$, under the influence of local fluctuating magnetic fields $\vec{h}(t)$.
In fact, $T_1^{-1}$ probes the spectral density $J(\omega_L)$ of the fluctuating field components $h_{\perp}(t)$ 
perpendicular to the applied magnetic field, 
at the Larmor frequency $\omega_L$:
\begin{equation}
\frac{1}{T_1} = \frac{\gamma^2}{2}\int_{-\infty}^{\infty} \langle h_{\perp}(t)h_{\perp}(t+\tau) \rangle_t\exp(-i\omega_L\tau) d\tau \, ,
\end{equation}
with $\gamma$ being the nuclear gyromagnetic ratio and $\langle h_{\perp}(t)h_{\perp}(t+\tau) \rangle_t$ being the autocorrelation function of the fluctuating magnetic field, which is assumed to decrease exponentially with the characteristic correlation time $\tau_c$: $\langle h_{\perp}(t)h_{\perp}(t+\tau) \rangle_t = \langle h_{\perp}^2\rangle \exp\left(-|\tau|/\tau_c\right)$. This leads to:
\begin{equation}
T_{1,BPP}^{-1}(T) = \gamma^2h_\perp ^2\frac{\tau_c(T)}{1+\tau_c^2(T)\omega_L^2} \, ,
\label{eq:T1BPP}
\end{equation}
and thus to a peak in $T_1^{-1}$ at the temperature where the effective correlation time of the spin fluctuations $\tau_c$ equals the inverse of the Larmor frequency $\omega_L$.
For a glassy spin freezing, the temperature dependence of the correlation time of the spin fluctuations can be described by an activated behavior:\cite{Suh2000,Curro2000,Curro2009}
\begin{equation}
\tau_c(T) = \tau_0\exp(E_a/k_BT) \, ,
\end{equation}
with the activation energy $E_a$ and the correlation time at infinite temperature $\tau_0$.
Thus, upon applying higher magnetic fields, the peak in \slrr\ (and correspondingly a peak in \slrrt) should shift to higher temperatures, which is what we observe experimentally.

\begin{figure}[t]
\centering
\includegraphics[width=\linewidth]{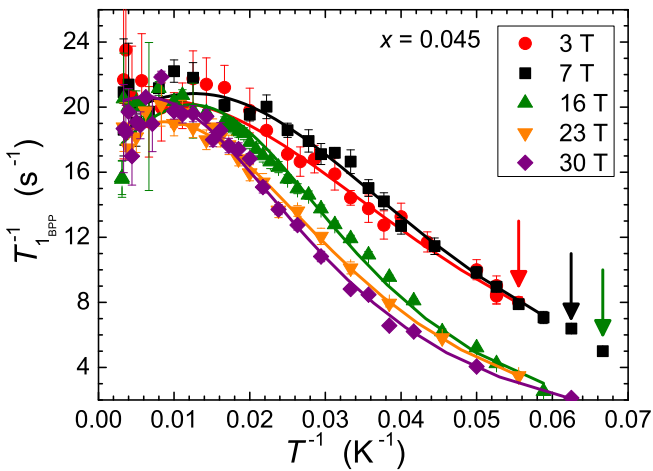}
\caption{\label{fig:Tinv}  (color online) $T_{1,BPP}^{-1}$ of LaFeAsO$\mathrm{_{0.955}}$F$\mathrm{_{0.045}}$ versus inverse temperature, calculated by subtracting the linear contribution of Eq.~\eqref{eq:BPP} from the measured \slrrt. Arrows mark $T_c^{-1}(H)$. Solid lines denote the BPP part of our fits, see Eq.~\eqref{eq:T1BPP}.}
\end{figure}

To analyze our data, we combine the BPP model for spin fluctuations with a linear temperature dependence of $(T_1T)^{-1}$, to account for the high temperature behavior, which apparently has another origin, since it is also visible in the optimally-doped and over-doped samples, where the peak in $(T_1T)^{-1}$ has disappeared. In the end our fitting function reads:
\begin{equation}
(T_1T)^{-1} = a + bT+\left(\frac{1}{T}\right)T_{1,BPP}^{-1} \,.
\label{eq:BPP}
\end{equation}

The constants $a$ and $b$ describe the linear temperature dependence. The last part is the BPP model (see Eq.~\eqref{eq:T1BPP}), multiplied with a prefactor $(1/T)$ to account for the fact that we actually fit the $(T_1T)^{-1}$ data.
This formula is used to fit the data of the underdoped samples ($x=0.045/0.05/0.075$) in 7\,T over the whole temperature range down to the onset of superconductivity at $T_c$, by fixing the slope $b$ of the linear temperature dependence to the value found in the overdoped sample with $x=0.15$. The constant term $a$ was found to vary only slightly between the three fits. The data of $x=0.1$ and $x=0.15$ were only fitted with the linear contribution. All corresponding fitting curves are plotted in Fig.~\ref{fig:dop} and (in an enlarged scale) in Fig.~\ref{fig:zoom} as solid lines. Note that if the linear fits were only applied to the high temperature points for $x=0.1$, the resulting slope $b$ is the same as that of the $x=0.15$ sample (shown as dashed lines in Figs.~\ref{fig:dop} and~\ref{fig:zoom}). Thus, the deviation of the data from the linear dependence at low temperatures suggests that remnants of spin fluctuations remain even for this optimally-doped sample.

The fact that a doping-independent slope $b$ can be used to describe the linear temperature dependence of the dynamic susceptibility at high temperatures, as measured by \slrrt, agrees very well with the observed doping-independent high-T linear
slope of the macroscopic susceptibility.\cite{Klingeler2010}
These observations point towards a ground-state independent origin, such as the
recently-suggested density of states effects.\cite{Skornyakov2011} 
Note that setting $b$ as a free parameter during the fitting procedure did not significantly change
the resulting BPP fit parameters.
The doping-dependence of the BPP fit parameters $E_a$,
$h_\perp$ and $\tau_0$ is plotted in a phase diagram of spin
fluctuations in Fig.~\ref{fig:parameter}a)
together with the corresponding superconducting transition
temperatures. Their absolute values and evolution upon doping will
be discussed in Section~\ref{discuss}.

\begin{figure}[t]
\centering
\includegraphics[width=\linewidth]{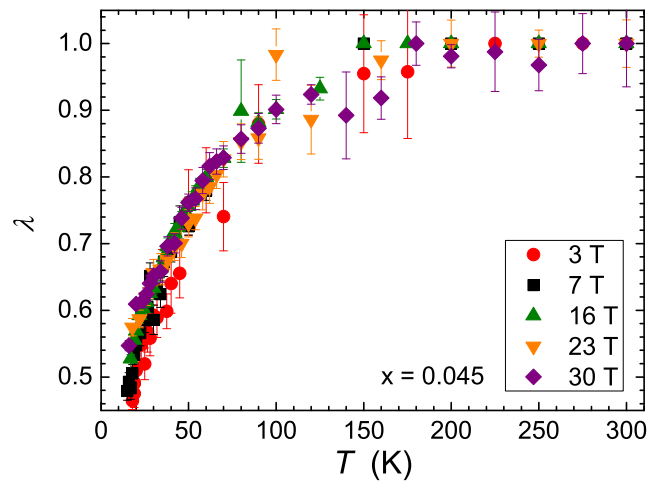}
\caption{\label{fig:lambda} (color online) Temperature dependence of the stretching exponent $\lambda$ of Eq.~\eqref{eq:T1} for the field-dependent measurements on LaFeAsO$\mathrm{_{0.955}}$F$\mathrm{_{0.045}}$.}
\end{figure}

Also the field-dependent data of the sample with $x$ = 0.045 can be well fit with the BPP-like model introduced in
Eq.~\eqref{eq:BPP}. Here again we fixed  
$b$ to the
value found for $x=0.15$ in $H_0=7$\,T. The fits are shown in
Fig.~\ref{fig:field} as solid lines and the corresponding BPP
parameters are collected in
Fig.~\ref{fig:parameter}b) and analyzed in Section~\ref{discuss}.
To isolate and illustrate the BPP contribution, Fig.~\ref{fig:Tinv} represents the field-dependent $T_{1,BPP}^{-1}$ data of $x=0.045$ versus inverse temperature, the usual
representation of the BPP-model,\cite{Heitjans2005, Suh2000} after
having subtracted the linear contribution from \slrrt.
Solid lines are the corresponding BPP-parts of our fits (see
Eq.~(\eqref{eq:T1BPP})). A clear field dependence is visible and
corroborates our choice of the BPP model.

Fig.~\ref{fig:lambda} shows the temperature evolution of the stretching exponent $\lambda$ used to fit the recovery of the nuclear magnetization (see Eq.~\eqref{eq:T1}) for the field-dependent measurements on the sample with $x$ = 0.045. A similar evolution was found for the doping-dependent measurements on all underdoped samples in 7\,T (not shown). Starting from the temperature where \slrrt\ begins to increase, a stretching exponent $\lambda < 1$ had to be used. This behavior, pointing towards a distribution of spin-lattice relaxation rates around a characteristic \slrr, is in good agreement with the suggested slowing down of spin fluctuations in this temperature range resulting in a glassy spin freezing.\cite{Curro2000, Julien2001} Note that in principle this distribution of spin-lattice relaxation rates could also stem from the presence of two different electronic environments on the nanoscale as observed by $^{75}$As NQR.\cite{LangPRL2010} However, these nanoscale regions are already present at room temperature, where the recovery of the nuclear magnetization is still well describable with $\lambda = 1$.
\begin{figure}[t]
\centering
\includegraphics[width=\linewidth]{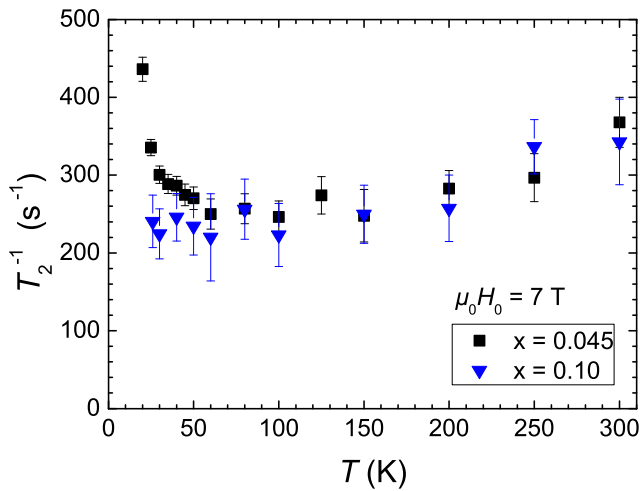}
\caption{\label{fig:parameter} (color online) Spin-spin relaxation rate, $T_2^{-1}$, versus temperature, measured in a magnetic field of 7\,T for $x=0.045$ (black squares) and $x=0.1$ (blue triangles).}
\label{fig:T2}
\end{figure}

A further manifestation of the progressive slowdown of spin
fluctuations in the underdoped samples is shown in Fig.~\ref{fig:T2}, which compares
the spin-spin relaxation rate $T_2^{-1}$ of an underdoped sample
($x=0.045$) with the one of the optimally-doped sample
($x=0.1$), both obtained in an external magnetic field of 7\,T by
fitting the decay of the spin echo after a $\frac{\pi}{2} - t -
\pi$ pulse sequence to:
\begin{equation}
M_{xy}(2t)=M_0\rm{e}^{(-2t/T_2)} \, . \label{T2fit}
\end{equation}
While the spin-spin relaxation rate of the optimally-doped sample decreases with decreasing temperature and levels off at
a roughly constant value for $T\leq 100$\,K, $T_2^{-1}$ of the
underdoped sample shows the same decrease at high
temperatures, but increases below $T\approx 60$\,K. Connected with these observations is
a broadening of the NMR linewidth for $x=0.045$ with decreasing
temperature, which is absent for $x=0.1$. However, this broadening
is difficult to quantify since the measurements for $x=0.045$ have
been done on a powder sample which also prevents a detailed
analysis of $T_2$, such as a subtraction of the Redfield contribution, stemming from spin-lattice relaxation processes. Nevertheless, the values of $T_2$ for $T$ $>$ 50\,K are of the same order of magnitude as those found by Oh \textit{et al.} in \bfca,
\cite{Oh2011} where they were argued to reflect a a spin-motional narrowing.
The upturn of $T_2^{-1}$ at low temperatures for $x=0.045$ could either be caused directly by increased electronic spin fluctuations, or from a reduction in the indirect interaction leading to spin-motional narrowing. On the basis of our measurements on powder samples we cannot distinguish between these two effects. Nevertheless, the lack of such an increase for $x=0.1$ is consistent with our previous interpretation of the \slrrt\ data. The fact that the increase of $T_2^{-1}$ starts at a different temperature than the one of \slrrt\ suggests a slowing down of spin fluctuations over a broad temperature range, i.e., the lack of a well-defined transition, and thus refutes a scenario where the increase of \slrrt\ is caused by a simple magnetic ordering.

\begin{figure}[t]
\centering
\includegraphics[width=\linewidth]{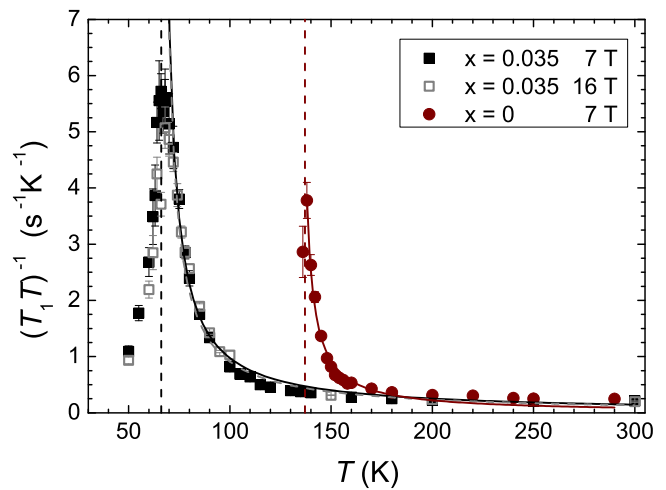}
\caption{\label{fig:mag} (color online) \slrrt\ in the undoped (dots) and underdoped
($x=0.035$, squares) samples, which become magnetic below $T_N=137$\,K and
$T_N=65$\,K (dashed lines), respectively. Solid lines are Curie-Weiss-Fits to the data. For $x=0.035$, measurements in 7\,T (filled black squares) and at 16\,T (open grey squares) are shown. No field dependence could be
resolved.}
\end{figure}

As a final illustration of the specific character of the relaxation upturn seen
in the underdoped superconducting samples, relaxation data for two magnetically-ordered samples ($x=0$ and $x=0.035$) at different fields (7\,T and 16\,T)
are shown in Fig.~\ref{fig:mag}. $(T_1T)^{-1}$ for these
samples increases strongly towards the corresponding magnetic
ordering temperatures. The absolute values of \slrrt\ at the
maximum are about an order of magnitude higher than in the
underdoped, superconducting samples. The data can be
well fit with a simple Curie-Weiss law $(T_1T)^{-1} =
C/(T+\theta)$ where $\theta = -T_N$ yields the magnetic ordering
temperature $T_N$.\cite{Ning2010, Oka2012, Kinouchi2011} For the undoped sample we find $T_N = 137$\,K,
in nice agreement with earlier macroscopic susceptibility and muon spin rotation ($\mu$SR) measurements.\cite{Klingeler2010,LuetkensNatMat2009} For the sample with $x=0.035$ we find $T_N=65$\,K, which compares well
with $T_N = 58$\,K for a sample with a nominal fluorine content of
$x=0.03$, as recently reported by $^{75}$As NQR measurements.\cite{Oka2012} 

The most essential point of the measurements is the fact that, in contrast to the peak observed in $(T_1T)^{-1}$ in the underdoped superconducting
samples, the peak in $(T_1T)^{-1}$ of $x=0.035$ does not shift
with field. The measurements in 16\,T yield the same temperature
dependence of \slrrt\ and a fit to the data results in the same
magnetic ordering temperature, $T_N=65$\,K. This behavior confirms that the
peak in \slrrt\ in the underdoped superconducting samples does not
stem from a weak magnetic ordering, as suggested before,\cite{NakaiNJP2009}
 but is indeed related to another effect, which we
identify as a glassy spin freezing.

\section{Discussion}
\label{discuss} 

\begin{figure}
\centering
\includegraphics[width=\linewidth]{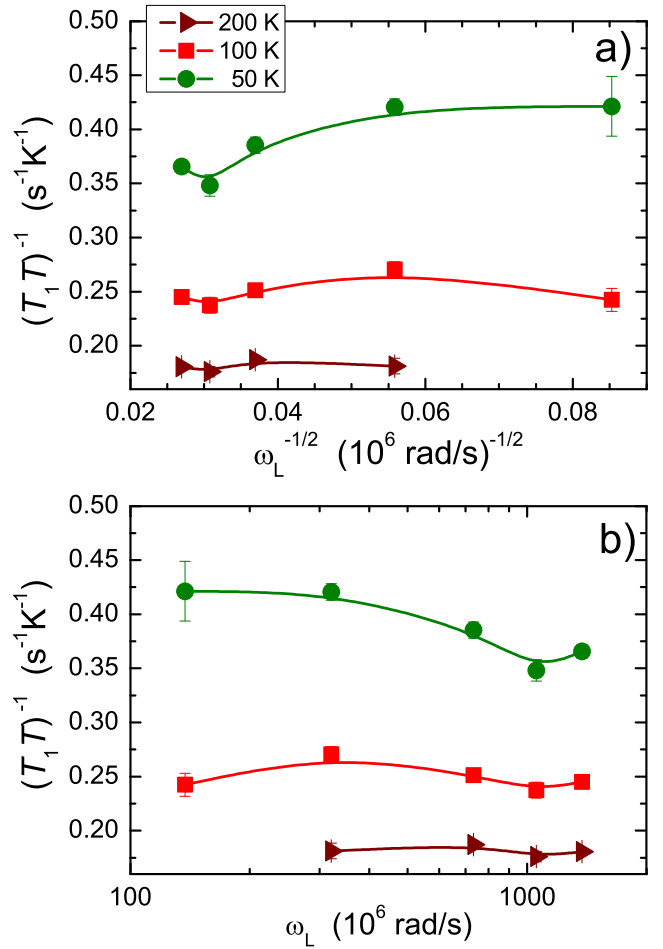}
\caption{\label{fig:spindiffusion} (color online) Tests for possible spin diffusion mechanism in (a) one and (b) two dimensions at three selected temperatures. (a) \slrrt\ versus $\omega_L^{-1/2}$, which should result in a linear dependence if 1D spin diffusion effects are at play. (b) \slrrt\ versus $\omega_L$ on a semilogarithmic scale, testing for $(T_1T)^{-1}\propto \ln\omega_L^{-1}$ which is expected if 2D spin diffusion effects are at play. Lines are guides to the eyes.}
\end{figure}

Before discussing the doping and field dependence of the resulting BPP fitting parameters $E_a$, $h_\perp$ and $\tau_0$, let us consider other effects than the BPP mechanism, which could also lead to a field-dependent peak in \slrrt, but which could be ruled out for the present case. One such effect would be that the application of a magnetic field introduces some field-induced anisotropy in the spin fluctuations and thus a field-dependence.\cite{Suh1995} This is in disagreement with our measurements on the sample with $x=0.035$, which in this case should exhibit some field dependence, as observed for $x=0.045$. Another possible effect to be taken into consideration is electronic spin diffusion in low dimensions.\cite{Julien2008} In fact, for electronic spin diffusion in one dimension, one expects a field dependence of the spin-lattice relaxation rate of the form $(T_1T)^{-1}\propto \omega_L^{-1/2}$, while for 2D spin diffusion a dependence of the form $(T_1T)^{-1}\propto \ln\omega_L^{-1}$ is expected.\cite{Julien2008, Kambe1994} 
For our field-dependent measurements in magnetic fields up to 30\,T we do not find any of these two dependencies (see Fig.~\ref{fig:spindiffusion}), so that we can also exclude spin diffusion effects as the source of the observed field dependence of \slrrt.  Finally, since the maximum of \slrrt\ shifts to higher temperature upon increasing the magnetic field, a (pseudo) spin gap scenario such as discussed in the cuprates\cite{Berthier1996} appears also very unlikely. 

Let us now turn to the discussion of the obtained BPP fitting parameters.
Fig.~\ref{fig:parameter} shows $E_a$, $\tau_0$ and $h_\perp$ as a function of doping
as deduced from measurements in 7\,T (Fig.~\ref{fig:parameter}a)) and as a
function of magnetic field as deduced from measurements on the
sample with $x=0.045$ (Fig.~\ref{fig:parameter}b)). We first concentrate on the
doping dependence (Fig.~\ref{fig:parameter}a)). The correlation time at infinite temperature and the value of the
fluctuating magnetic field are essentially doping-independent and
amount to $\tau_0=2.3$\,ns and $h_\perp =25$\,Oe. The rather long $\tau_0$ is comparable to the value found in stripe-ordered
La$_{1.65}$Eu$_{0.2}$Sr$_{0.15}$CuO$_4$, where Simovi\v{c}
\textit{et al.} found $\tau_0 = 1.3$\,ns by fitting the \la\
\slrr\ to the standard BPP model.\cite{Simovic2003}
The absolute value of $h_\perp$ could be consistent with ZF-$\mu$SR and M\"ossbauer studies, where indications of low-T static disordered magnetism were found in underdoped samples with $0.05
\leq x \leq 0.075$, with internal fields being a factor 20 smaller than in the magnetically-ordered
sample with $x=0.04$, where $H_{int} \approx$ 1600\,Oe.\cite{Luetkens2008, LuetkensNatMat2009} This leads to a rough estimation of the internal fields in these underdoped samples of about $80$\,Oe,
which is of the same order of magnitude as our deduced $h_\perp$.
Also LF-$\mu$SR measurements on a superconducting sample with a nominal fluorine content of $x$ = 0.06 found evidence for very slowly fluctuating local spins and a spin-glass-like magnetic phase in 25$\%$ of the sample,\cite{Takeshita2008} while no sign of such slow spin fluctuations was observed in optimally- and over-doped \LaOFxFeAs.\cite{Ohishi2011} However, while the general observation of very slow spin fluctuations in underdoped samples by means of $\mu$SR is consistent with our findings, the fact that the magnetism in the underdoped samples as seen by ZF- and LF-$\mu$SR is diluted or occurs only in a minor sample volume, does not agree well with our finding of a glass-like transition of the spin fluctuations.
In this context, longitudinal field (LF) $\mu$SR on our samples could be of interest.

The activation energy $E_a$ increases with increasing
doping, from 33\,K for $x=0.045$ to 52\,K for $x=0.075$. A priori this would mean that the (thermal) energy that is needed to overcome
the spin freezing increases when moving away from the magnetically-ordered ground state, i.e., in opposite
to what one would naively expect. In the stripe-ordered cuprates,
such a doping dependence can be observed around 1/8$th$ doping,
where the temperature of the peak, and thus the activation energy,
is maximal.\cite{Grafe2010} However, beyond the comparable order of magnitude of $E_a$ in \LaOFxFeAs\ and in
such cuprates,\cite{Simovic2003,Curro2000} there is no ground to advocate such a scenario in pnictides. A more detailed investigation of the doping dependence may clarify this question.

\begin{figure}
\centering
\includegraphics[width=\linewidth]{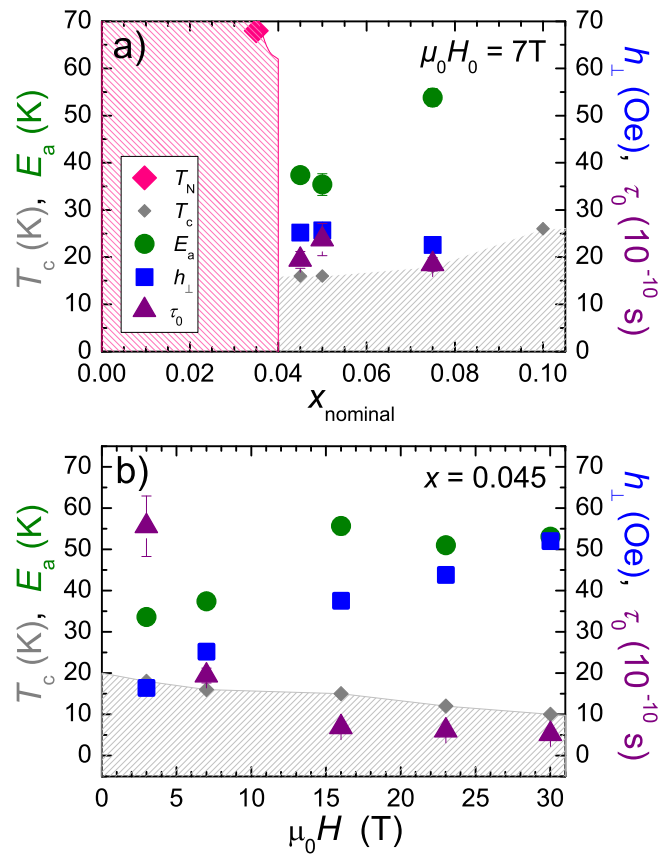}
\caption{\label{fig:parameter} (color online) Fit parameters to our BPP-like model (Eq.~\eqref{eq:BPP}) versus doping (a), obtained from the measurements in 7\,T, and versus field (b), obtained from the measurements on $x=0.045$. Grey shaded areas denote the superconducting ground state with grey diamonds marking $T_c(H)$. The pink-shaded area in (a) denotes the magnetically-ordered ground state, with the pink diamond indicating $T_N = 65$\,K for $x=0.035$ as deduced from \slrrt\ (see Fig.~\ref{fig:mag}).}
\end{figure}

We now turn to the field-dependence of the BPP fit parameters for
$x=0.045$ (see Fig.~\ref{fig:parameter}b)). An
increase of the fluctuating field $h_\perp$ and of the activation energy $E_a$ upon increasing the magnetic field, together with a decrease of the correlation time of the spin fluctuations at infinite temperature, $\tau_0$, with increasing
magnetic field suggest that spin fluctuations are enhanced upon applying an external magnetic field. This
observation is in agreement with transverse field (TF)-$\mu$SR measurements on
underdoped samples in the superconducting state, which showed an increase of magnetic correlations upon increasing the applied magnetic field.\cite{Luetkens2008} Note that, while the opposite field dependencies of $T_{max}$ and $T_c$ (see Table~\ref{tab:Tc4.5}) could suggest a competition between superconductivity and magnetic correlations, this could be explained by a direct reinforcement of the magnetism by the applied field, whereas the concomitant increase of $T_{max}$/$E_a$ and $T_c$ with doping would rather indicate that superconductivity may be intimately related to very slow spin fluctuations. Let us finally note that the fact that the magnetic energy of 16 Tesla is rather close to $E_a$
may lead to a field dependence of the BPP parameter that is
typically not observed.

Our data and interpretation
agree well with resistivity measurements on underdoped
\LaOFxFeAs\ where at temperatures below $\sim$60\,K an upturn of
the resistivity has been observed.\cite{Hess2009} This upturn is
indicative of charge carrier localization, and is visible in the
underdoped superconducting samples up to $x=0.075$.
These observations suggest that a remnant feature of the spin density wave (SDW) order
is still influencing the physics of the underdoped compounds,
despite the absence of long range magnetic order.

Note that our investigations are also in line with a recent NMR investigation 
on a sample with a nominal fluorine content of $x=0.04$.\cite{Nakai2012} In this sample, static magnetism was observed at $T_N =30$\,K, along with a greatly-reduced static magnetic field of $\mu_0H < 100$\,Oe, while a superconducting transition was observed at $T_c = 16.2$\,K.\cite{Nakai2012} The very small value of the internal magnetic field, which was actually deduced from a line broadening (and not from a clear splitting) of the $^{139}$La NMR spectrum, together with the reduced value of \slrrt\ observed in this sample suggest that our slow spin freezing scenario could bring insight on its physics. This sample seems to fit perfectly between our magnetically-ordered sample with $x=0.035$ and our superconducting sample with slow spin fluctuations with $x=0.045$.

Finally, note that on the basis of the presented measurements, the true nature of the observed slowed-down spin fluctuations cannot be revealed. Apart from antiferromagnetically-correlated spins, also charge, stripe or domain wall fluctuations could give rise to the enhancement in \slrrt. Further investigations are needed to clarify this point.

\section{Conclusion}

We have used $^{75}$As NMR to investigate different samples of
fluorine-doped \LaOFxFeAs. For underdoped samples with $0.045\leq
x\leq 0.075$, where the optimal $T_c$ is still not reached, we
find an enhancement of \slrrt\ with decreasing temperature,
followed by the formation of a well-defined peak in \slrrt\
clearly above the superconducting transition temperature. We
investigated the field-dependence of this peak in magnetic fields up to 30\,T and found a shift
of the maximum of \slrrt\ towards higher temperatures upon
increasing the field. This behavior is consistent with the BPP
model, describing a progressive slowing down of spin fluctuations. We
fit our doping- and field-dependent data with a model combining
the BPP-dependence of \slrr\ with a linear temperature contribution
to \slrrt, which is observed for the optimally- and over-doped
samples ($x=0.1$ and $x=0.15$) and seems to be doping-independent, in nice
agreement with the doping-independent slope of the macroscopic
susceptibility. The combination of these two contributions is able
to describe the doping- and field-dependent \slrrt\ data
of the underdoped samples over the whole temperature
range, from 480\,K down to $T_c$. Our model suggests the presence of very slow spin dynamics 
in underdoped, superconducting samples, which fully disappear beyond optimal doping. 
The field-dependence of the BPP fitting parameters suggests that spin fluctuations are enhanced upon applying an
external magnetic field. While this observation could suggest a competition between superconductivity and magnetic correlations, the doping-dependence of the peak in \slrrt\ and of the activation energy $E_a$ would rather indicate that superconductivity may be intimately related to very slow spin fluctuations. In this regard, further investigating the nature of these fluctuations would be of interest, together with their connection or lack thereof with the magnetism of the parent compound.

\section*{Acknowledgements}

The authors thank C.~Hess, S.~Sanna, and N.~Curro for valuable discussion and M. Deutschmann, J. Werner,
and R. Vogel for technical support. This work has been supported
by the Deutsche Forschungsgemeinschaft (DFG) through SPP1458 (Grant No.
GR3330/2, BE1749/13 and WU595/3-1). A portion of this work was performed at the National High Magnetic Field Laboratory, which is supported by National Science Foundation Cooperative Agreement No. DMR-1157490, the State of Florida, and the U.S. Department of Energy.


\end{document}